\documentclass[
reprint, 
 superscriptaddress,
 longbibliography,
nofootinbib,
nobibnotes,
 aps,
prl,
]{revtex4-2}

\usepackage[breaklinks=true,colorlinks,citecolor=blue,linkcolor=blue,urlcolor=blue]{hyperref}

\usepackage{tikz,xcolor,hyperref}
\usepackage{graphicx}
\usepackage{dcolumn}
\usepackage{amsmath}
\usepackage[T1]{fontenc}       
\usepackage{amssymb}
\usepackage{bm}
\usepackage{xspace}
\usepackage[utf8]{inputenc}
\usepackage{makecell}
\usepackage[T1]{fontenc}
\usepackage{etoolbox}
\usepackage{txfonts}
\usepackage{xcolor}
\usepackage{booktabs}
\usepackage{fixltx2e}
\usepackage{ragged2e}
\usepackage{url}

\makeatletter
\def\@email#1#2{%
 \endgroup
 \patchcmd{\titleblock@produce}
  {\frontmatter@RRAPformat}
  {\frontmatter@RRAPformat{\produce@RRAP{*#1\href{mailto:#2}{#2}}}\frontmatter@RRAPformat}
  {}{}
}%
\makeatother
\begin{document}

\title{Real-Time Relaxation Dynamics of Dark Excitons in Mirror-Symmetric Monolayers}
\author{Rishabh Saraswat}
\affiliation{Nanoscale Electro-thermal Laboratory, Indian Institute of Information Technology Allahabad, Uttar Pradesh 211015, India}
\author{Mantasha Ansari}
\affiliation{Electronic Structure Theory Group, Department of Electronics and Communication Engineering, Indian Institute of Information Technology Allahabad, Uttar Pradesh 211015, India}
\author{Rekha Verma}
\affiliation{Nanoscale Electro-thermal Laboratory, Indian Institute of Information Technology Allahabad, Uttar Pradesh 211015, India}
\author{Sitangshu Bhattacharya}
\thanks{Contact author: sitangshu@iiita.ac.in}
\affiliation{Electronic Structure Theory Group, Department of Electronics and Communication Engineering, Indian Institute of Information Technology Allahabad, Uttar Pradesh 211015, India}

\begin{abstract}
\noindent Symmetry governs which exciton-phonon (\textit{exc-ph}) scattering channels \textit{can} exist, but not which a relaxing exciton actually uses. Combining group theory with \textit{ab initio} many-body perturbation theory, we derive the mirror parity $\sigma_h$, of finite-momentum excitons in planar $D_{3h}$ monolayers, and form exact \textit{exc-ph} selection rules that hold across the whole Brillouin zone. The parity of the lowest exciton band changes across a closed contour. The even phonons connect states of the same parity, while odd phonons connect states of opposite parity. We use these rules to compute exciton linewidths and we solve the real-time Boltzmann transport equation for monolayer GaN as the test case and benchmark it against monolayer \textit{h}BN. 
\end{abstract}
\maketitle

\noindent \textit{Introduction---} 
Reduced dielectric screening binds excitons in atomically thin
semiconductors by hundreds of meV~\cite{Chernikov2014,He2014,Qiu2016,Wang2018}.
Phonons, however, rapidly redistribute these excitons across the
Brillouin zone (BZ)~\cite{Cassabois_2016, Rosati2026}, transferring population into
finite-momentum states that are dark to direct optical
recombination. Whether the momentum-dark excitons then luminesce at all depends on which phonons can scatter them~\cite{Cassabois_2016,Xiao_2015,Malic2018,Tang2019,Li2019,Rosati2020Cascade,Brem2020}. For such a dark state, phonon-assisted emission is the only way out, and the resulting indirect photo-(P)~\cite{Arnaud2006, Funk2021} and cathodo-(C)~\cite{Schue2019} luminescence emission (LE) spectrum is a direct readout of the exciton--phonon (\textit{exc-ph}) coupling. Group theory~\cite{Dresselhaus2008} has long been used with recently developed \textit{ab-initio} techniques~\cite{Galvani2016, Chen2020, Chen2022, nalabothula2025,Bajaj2025, Saraswat2025} to sort these processes into allowed and forbidden, thereby labelling these finite momentum excitons by their irreducible representations. Yet a selection rule states only what may happen, not what does. 
\begin{figure}[!h]
  \centering
  \includegraphics[width=\columnwidth]{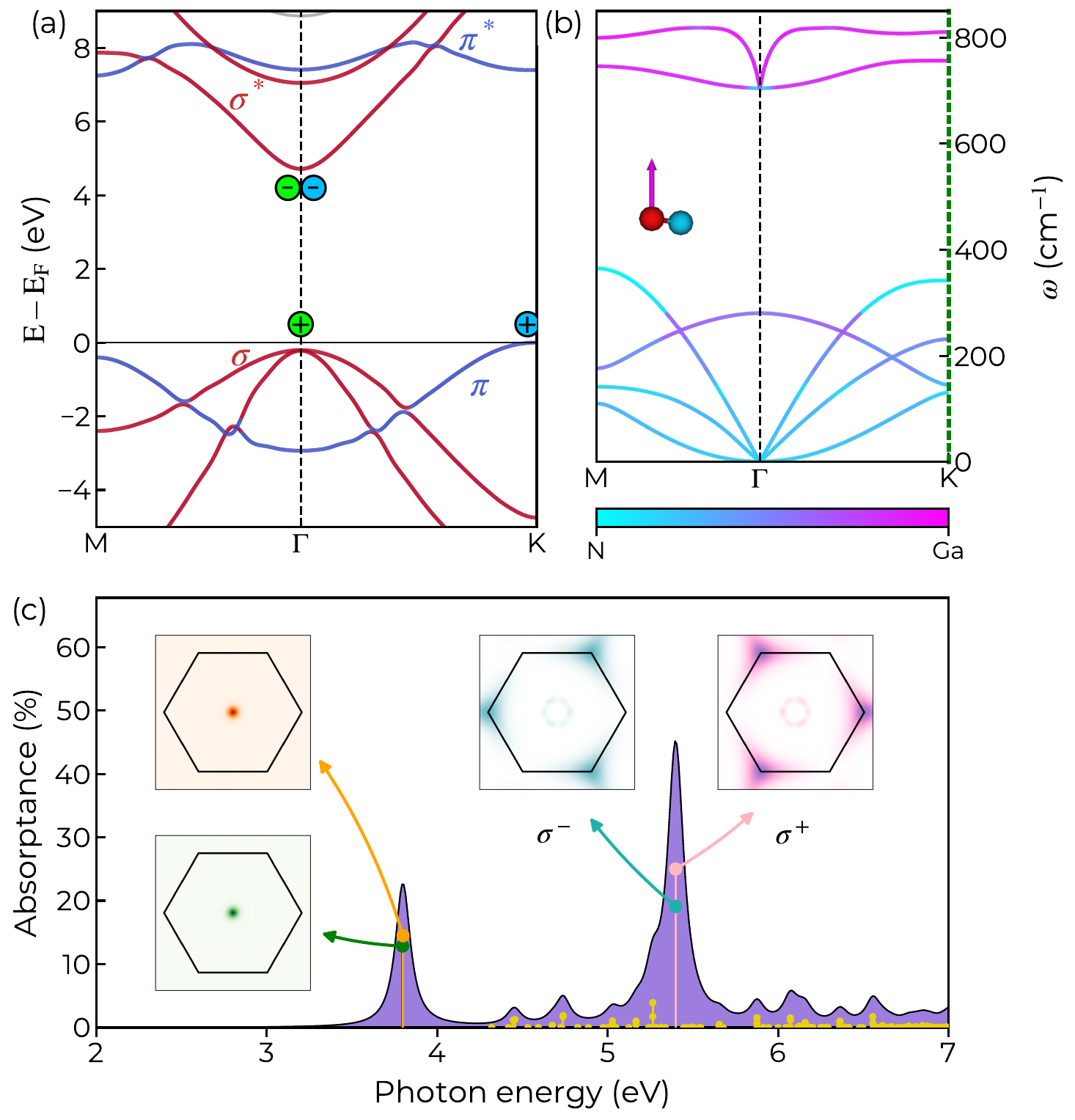}
  \caption{\justifying (a) G$_0$W$_0$ band structure of monolayer GaN with band symmetries, with parities of the bands overlaid. Blue is $p=-1$, while red is $p=+1$, with the $\sigma$ and $\pi$-like band edges labelled. Lime and blue markers denote the bright ($\mathbf{Q}=\mathbf{0}$) and momentum-dark excitons, respectively. (b) Phonon dispersion with the green dashed line marks modes providing the requisite momentum for phonon-assisted PLE. Inset: atomic displacements of the lowest energy mode (Ga: red, N: blue). The colorbar defines each atoms contribution. (c) Absorptance spectrum with excitonic oscillator strengths (yellow bars). The lowest-energy and brightest doubly degenerate excitons are highlighted in orange/green and teal/pink, respectively. Inset: Exciton de-polarization in the presence of left ($\sigma^-$) and right ($\sigma^+$) circularly polarized light.}
  \label{fgr:Fig_1}
\end{figure}
Which channels actually carry the population depends on where the excitons are, and this is exposed only by following their evolution in time, for example with the recent development of real-time Boltzmann transport equation (\textit{rt}-BTE)~\cite{Rosati2020Cascade,Chen2022, Rosati2026}. Whether parity rules can also decide relaxation and indirect emission across the whole Brillouin zone (BZ) remains an open question.\\
\noindent In this Letter, we answer it using monolayers with a horizontal mirror plane $\sigma_h$, a symmetry that assigns an even or odd parity to every state. We combine first-principles \textit{exc-ph} linewidths, an \textit{rt}-BTE solver developed for this work~\cite{Supplemental}, and a symmetry analysis that tracks the $\sigma_h$ parity of every exciton band across the BZ. With these, we ask (i) whether mirror parity merely forbids channels or also selects those that carry the relaxation, (ii) whether the parity of the band edges alone predicts how a planar monolayer emits light, (iii) which phonons release a dark lowest exciton radiatively, and whether symmetry fixes the polarization of the emitted light, and (iv) whether the same phonons set the linewidths and drive the relaxation.\\
\noindent The reason $\sigma_h$ is decisive is geometric. In monolayer semiconductors whose point group contains $\sigma_h$, such as $D_{3h}$ in \textit{h}BN and 1H transition-metal dichalcogenides (TMDCs), $\sigma_h$ maps every in-plane wavevector onto itself. It therefore belongs to the little group of every wavevector, and every electronic $|n\mathbf{k}\rangle$, phonon $|\nu\mathbf{q}\rangle$, and excitonic $|\lambda\mathbf{Q}\rangle$ state carries a definite mirror parity throughout the BZ. The resulting selection rules are exact: \textit{even} phonons connect excitons of equal parity, and \textit{odd} phonons connect excitons of opposite parity. Even excitons couple only to in-plane light, and odd excitons only to light polarized out of the plane. Buckling or Janus asymmetry removes $\sigma_h$, and with it all of these constraints. The parity of the lowest excitons is set by the parity of the band edges. In \textit{h}BN, the band edges share the same orbital parity, and the bright excitons are even.\\
We demonstrate these rules as a test case for planar GaN monolayer, which inverts this \textit{h}BN pattern. For contrast, we treat \textit{h}BN with the same workflow: its
lowest exciton is bright and even at $\mathbf{\Gamma}$, so the same rules that make the GaN lowest exciton dark and valley-polarized make the \textit{h}BN one bright and unpolarized~\cite{Supplemental}. Ultrathin wurtzite films are predicted to flatten into graphitic sheets that quench their polar dipole~\cite{Freeman2006}, as observed also in ZnO~\cite{Tusche2007} and AlN~\cite{Tsipas2013}. Freestanding monolayer GaN likewise relaxes into a planar $D_{3h}$ structure~\cite{Zhuang2013,Singh2014}, which our calculations confirm. GaN grown so far by graphene encapsulation on SiC is buckled~\cite{AlBalushi2016,Sfuncia2023}; only the planar phase keeps $\sigma_h$ exact, and it is this phase we study. \\
\noindent \textit{Parities and symmetry---} Fig.~\ref{fgr:Fig_1}(a) shows this contrasting trend through its indirect quasi-particle G$_0$W$_0$ band structure (for details, see \cite{ComputationalDetails}\nocite{QuantumEspresso,Sangalli2019,Supplemental,Sohier2017LOTO,letzelphc,Paleari2025}), opening a band gap of 4.91~eV, demonstrating CBM at $\mathbf{\Gamma}$ and the VBM at $\mathbf{K}$. A $\sigma$-like Ga-$s$ conduction band lies below the $\pi^*$ band, opposite in parity to the valence-band maximum, and the gap is indirect. The lowest exciton is therefore dark on two counts: it lies at finite momentum \textbf{Q}, and it is $\sigma_h$-odd. At $\mathbf{\Gamma}$, the top valence state is a doublet of $E'$ symmetry and the conduction-band edge is a singlet of $A_1'$ symmetry. An exciton formed directly at $\mathbf{\Gamma}$ from these two states must then transform as $\Gamma_{\mathrm{exc}} = \Gamma_c \otimes \Gamma_v^{*} = A_1' \otimes E' = E'$, and our Bethe Salpeter equation (BSE) calculations indeed find this exciton as a degenerate pair. The optical activity follows from the dipole operator, which in D$_{3h}$ splits into an in-plane part $E'\{x,y\}$ and an out-of-plane part $A_2''\{z\}$. In-plane light can therefore excite only $E'$ excitons, and out-of-plane light only $A_2''$ excitons. The strongest absorption, however, does not come from $\mathbf{\Gamma}$ but from direct transitions at the $\textbf{K}$ valleys as shown in Fig.~\ref{fgr:Fig_1}(c). These excitons also appear at $\mathbf{Q}=\mathbf{0}$ as $E'$ doublets, but their valley origin gives them additional structure. The little group of $\textbf{K}$ is $C_{3h}$, which has no vertical mirror, so each band state at $\textbf{K}$ or $\textbf{K}'$ is nondegenerate and has a well-defined crystal angular momentum $m$ (mod 3), given by $C_3|X\rangle = e^{-2\pi i m/3}|X\rangle$~\cite{nalabothula2025}. Choosing the basis of each doublet so that $C_3$ is diagonal, separates it into an $m=+1$ and an $m=-1$ partner, one in each valley. Each partner is fully circularly polarized ($\rho=\pm1$) and responds only to $\sigma^{+}$ or $\sigma^{-}$ light, with its $\mathbf{k}$-space weight in $\textbf{K}^{+}$ or $\textbf{K}^{-}$, respectively.
\begin{figure}[!ht]
  \centering
  \includegraphics[width=1\columnwidth]{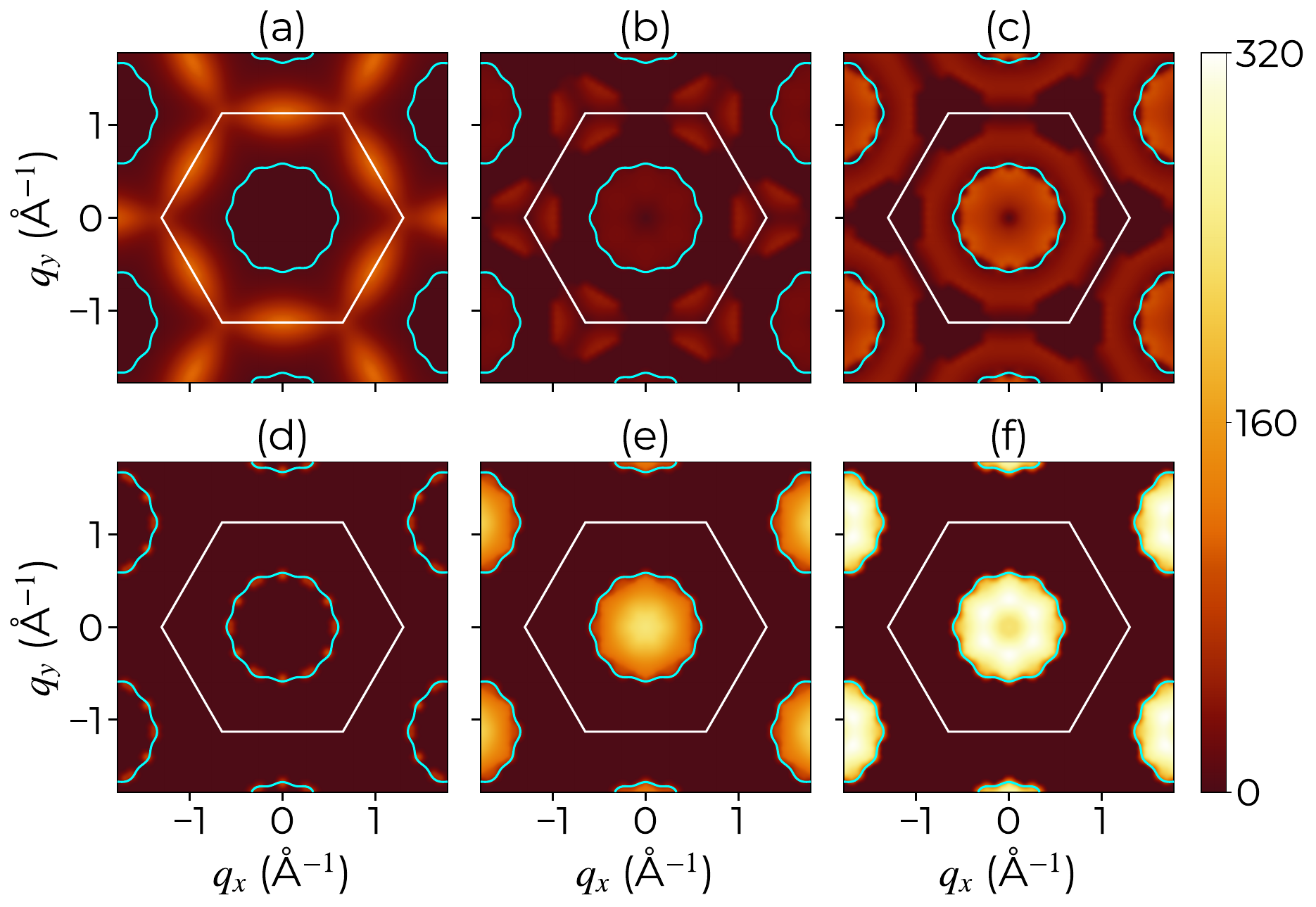}
  \caption{\justifying (a)-(f) The \textit{exc-ph} coupling strength, $\bigl|\mathcal{G}_{\beta\lambda\nu}(0,\mathbf{q})\bigr|^{2}$ (in 10$^2$~meV$^2$) from the lowest bright exciton at $\mathbf{\Gamma}$ ($\lambda=1$) to the lowest excitonic band ($\beta=1$) at $\mathbf{Q+q} = \mathbf{q}$, mapped across the BZ for the six energy-ordered phonon branches of monolayer GaN. The cyan line marks the transition of the lowest excitonic band from even parity ($p=+1$) to odd ($p=-1$). All panels share the same color scale.}
  \label{fgr:Fig_2}
\end{figure}

\begin{figure*}[ht]
  \centering
  \includegraphics[width=0.9\textwidth]{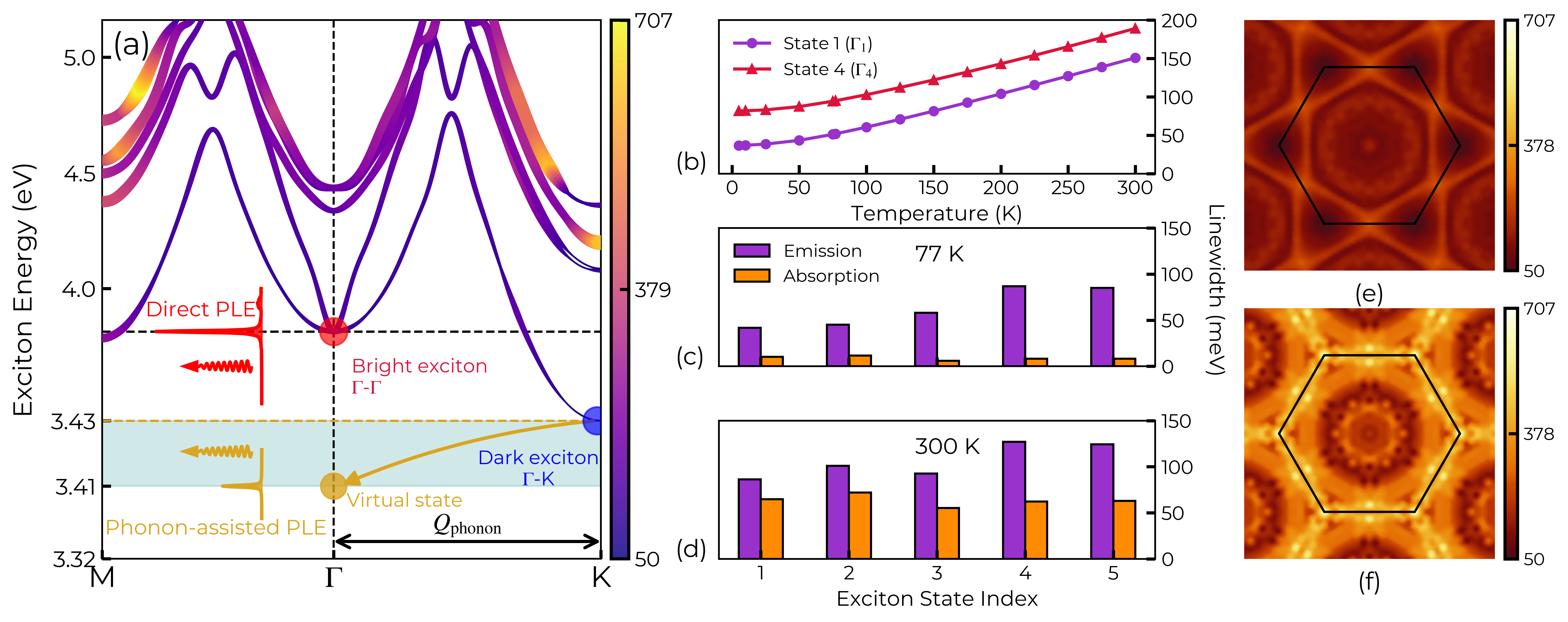}
  \caption{(a) Exciton dispersion for the five lowest excitons. The size and colormap of the bands represent the linewidth at 300~$\mathrm{K}$ for each point. The phonon-assisted PLE mechanism has been plotted against each band. The satellite is in gold, the bright peak is in red. $Q_{phonon}$ represents the phonon momentum that connects $\mathbf{K}$ to $\mathbf{\Gamma}$. (b) The linewidths for the exciton states 1 and 4 at $\mathbf{\Gamma}$ have been plotted for temperatures ranging from 5--300~$\mathrm{K}$. (c)-(d) The linewidths segregated by the emission and absorption components for the first 5 excitonic bands at the $\mathbf{\Gamma}$ point are plotted, at temperatures 77~K and 300~K respectively. (e)-(f) The linewidths at 300~$\mathrm{K}$ plotted for the states 1 and 4 across the BZ. (e) and (f) share one color scale.}
  \label{fgr:Fig_3}
\end{figure*}
\noindent Under $\sigma_h$, each electron--hole component of an exciton carries the parity $p_c(\mathbf{k}+\mathbf{Q})\,p_v(\mathbf{k})$, the product of the electron and hole parities at the wavevectors they occupy. The parity of the exciton is therefore a weighted sum of these products
\begin{equation}
\chi_\lambda(\mathbf{Q}) = \sum_{cv\mathbf{k}} \big|A^{\lambda\mathbf{Q}}_{cv\mathbf{k}}\big|^2\, p_c(\mathbf{k}+\mathbf{Q})\, p_v(\mathbf{k}),
\label{eq:parity}
\end{equation}
\noindent which we derive in the Supplemental Material (SM)~\cite{Supplemental} from the transformation of the BSE eigenvectors under a symmetry operation. Because $\sigma_h$ commutes with the BSE Hamiltonian at every $\mathbf{Q}$, all components with nonzero weight share the same product, and each exciton band is strictly even or odd. The parity of the bright excitons at $\mathbf{Q}=\mathbf{0}$ is therefore not inherited elsewhere in the zone; it must be determined at each $\mathbf{Q}$. Doing so reveals a global structure shown in Fig.~\ref{fgr:Fig_2}. The lowest exciton band, taken in energy order, is $\sigma_h$-even inside a closed contour around $\mathbf{\Gamma}$ and odd throughout the rest of the zone, including $\textbf{K}$ and $\textbf{M}$. The contour marks where an even and an odd band cross. The mirror symmetry forbids their mixing, so the crossing is protected and the parity of the lowest band switches across it. The momentum-dark exciton at $\mathbf{Q}=\textbf{K}$, the global minimum, lies in the odd region and transforms as $A''$ of the little group $C_{3h}$. It belongs to a band of opposite mirror parity from the bright $E'$ doublet.
\begin{table}[b]
\caption{\label{tab:modes}Phonon branches at $\mathbf{K}$ (from Fig.~\ref{fgr:Fig_1}(b)). Energies, irreps of $C_{3h}$ (C$_3$ taken about the N site at the origin), mirror parities and whether each can mediate the radiative decay of the $A^{\prime\prime}$ exciton under Eq.~(\ref{eq:pl}).}
\begin{ruledtabular}
\begin{tabular}{lccc}
$\hbar\omega$ (meV) & irrep & $\sigma_h$ & allowed \\
\hline
$16$  & $E_1''$ & odd  & \textbf{yes} \\
$18$  & $A''$ & odd  & no \\
$29$  & $E_2'$  & even & no \\
$42$  & $A'$  & even & no \\
$94$  & $E_1'$  & even & no \\
$100$ & $E_2'$  & even & no \\
\end{tabular}
\end{ruledtabular}
\end{table}

\noindent Mirror parity organizes the phonons in the same way. Phonon branch indices follow energy order and exchange character wherever two branches cross, so a branch index carries no fixed parity across the BZ. The parity of each mode, by contrast, is exact at every in-plane $\mathbf{q}$. The \textit{exc-ph} selection rules must therefore be applied mode by mode. $\sigma_h$ belongs to the little group of every $\mathbf{Q}$, hence, the \textit{exc-ph} matrix element obeys $\mathcal{G}_{\beta\lambda\nu}=p_\beta\,p_\nu\,p_\lambda\,\mathcal{G}_{\beta\lambda\nu}$ and vanishes unless $p_\beta p_\nu p_\lambda=+1$~\cite{Supplemental}. Even phonons connect excitons of equal parity, odd phonons excitons of opposite parity. Only an odd phonon can scatter the dark exciton into the bright $\mathbf{Q}=\mathbf{0}$ states. The maps in Fig.~\ref{fgr:Fig_2}(a)-(f) show scattering out of the lowest exciton at $\mathbf{Q}=\mathbf{0}$, which is $\sigma_h$-even. Every symmetry-forbidden channel vanishes to numerical precision across the entire BZ~\cite{Supplemental}, so the zeros in these maps are set by symmetry, not by weak coupling. The polar optical branches are even throughout the zone and couple only inside the even domain, while the out-of-plane acoustic (ZA) branch couples only outside it. The fourth branch is the instructive case. Near $\mathbf{\Gamma}$ it is the out-of-plane optical (ZO) mode, which is odd, but beyond a crossing at $q \approx 0.55$--$0.57$~\AA$^{-1}$ it becomes an even in-plane mode. It is therefore forbidden inside the even domain, where it is odd and the final state is even, and beyond ($q \approx 0.58$--$0.60$~\AA$^{-1}$) , where it is even and the final state is odd. It couples only in the annulus, and its map is a ring shown in Fig.~\ref{fgr:Fig_2}(d).\\
\noindent We symmetrized the \textit{exc}-\textit{ph} matrix elements by explicit rotation within each star of $\mathbf{q}$ and used our code to obtain the linewidths across all the excitonic BZ using linear-triangle integration method~\cite{Jepsen1971} on the energy-conserving delta functions on a mesh refined fifteen-fold ($360\times360$). Within each coarse cell, the three ingredients were interpolated separately. Exciton energies were taken from a local quadratic fit to the six nearest BSE points, following the hybrid tetrahedron scheme~\cite{Dong2025}; a global Fourier interpolation was avoided because it oscillates wherever energy-ordered exciton bands cross. The \textit{exc-ph} couplings $|G|^2$ were held constant within each coarse cell, except for the acoustic branches near $\mathbf{q}=0$, where $|G|^2 \propto q$ was fitted to the innermost shell. Convergence tests, the construction of the symmetry analysis, and details of the \textit{rt}-transport scheme are given in the SM~\cite{Supplemental}.\\
\noindent These parity selection rules determine how the dark exciton at $\mathbf{Q}=\textbf{K}$ decays radiatively. A phonon can mediate the emission only if
\begin{equation}\label{eq:pl}
  \Gamma_{\nu} \subset \Gamma_{\rm exc}^{*} \otimes \Gamma_{\rm photon},
\end{equation}
where $\Gamma_{\rm exc}$ is the representation of the initial exciton and $\Gamma_{\rm photon}$ that of the emitted photon, both taken in the little group $C_{3h}$ of $\textbf{K}$. Light emitted normal to the layer is polarized in-plane. Its representation $E'$ of D$_{3h}$ splits in C$_{3h}$ into the two one-dimensional irreps $E_1'$ and $E_2'$, one for each circular polarization. For the $A''$ dark exciton, Eq.~(\ref{eq:pl}) then requires a phonon of $E_1''$ or $E_2''$ symmetry. Only one of these, $E_1''$, occurs among the six phonon branches at $\textbf{K}$ (see Table~\ref{tab:modes}). Exactly one branch can therefore mediate the emission, and the emitted light is circularly polarized; by time reversal, the $\textbf{K}'$ valley emits the opposite helicity (See Fig.~S9 in SM~\cite{Supplemental}). Out-of-plane emission would require an $A'$ phonon, which contributes negligibly. 
\begin{figure*}[ht]
  \centering
  \includegraphics[width=0.9\textwidth]{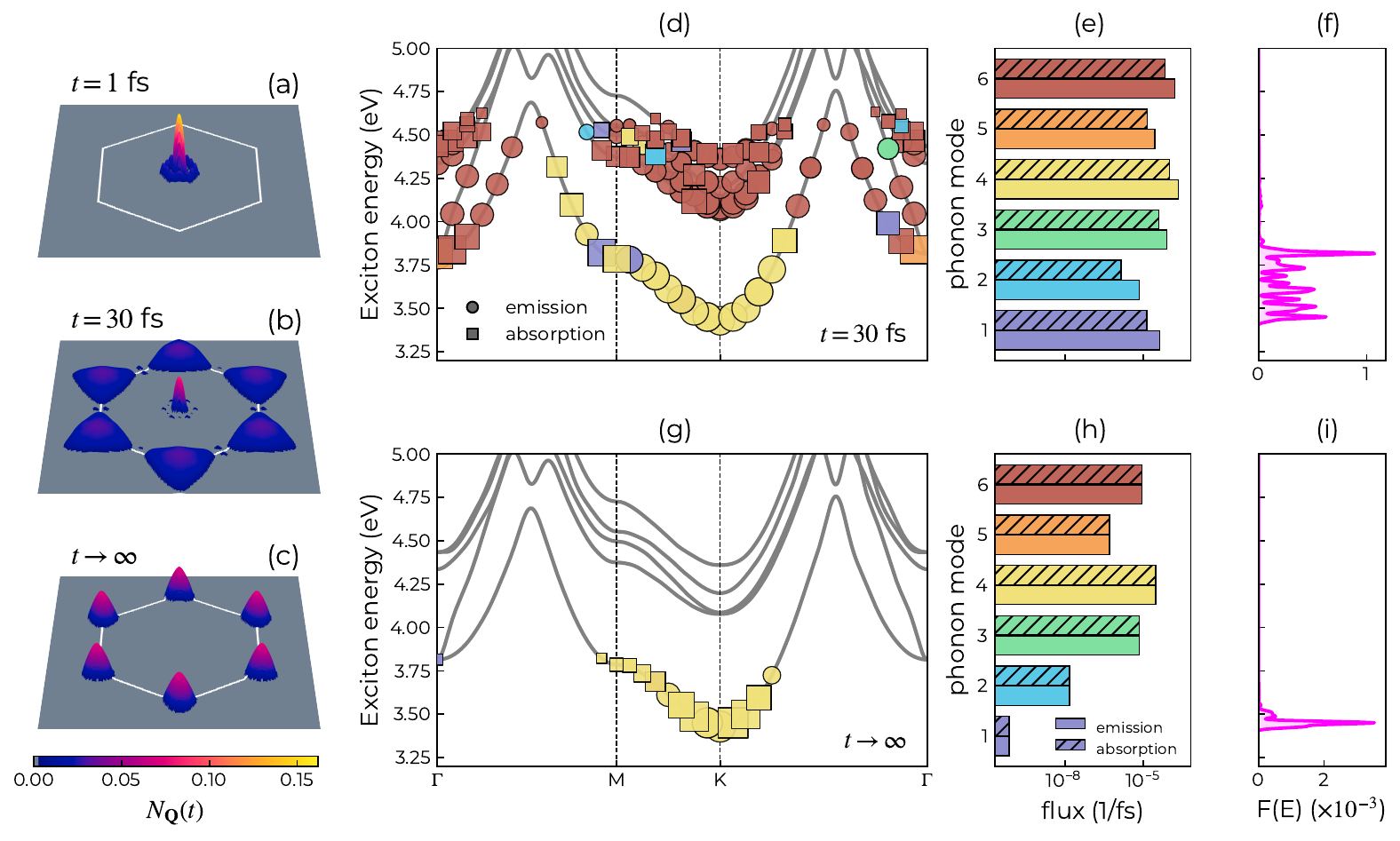}
  \caption{\justifying Real-time snapshots of exciton thermalization in monolayer GaN at $300$~K, following optical injection at $\mathbf{Q=0}$ weighted by oscillator strengths. (a)–(c) Exciton occupation $N_{\mathbf{Q}}$ across the Brillouin zone at $t = 1$~fs, $30$~fs, and after thermalization. (d),(g) The same population projected onto the exciton dispersion at $30$~fs and $t \to \infty$; marker area scales with occupation of the dominant channel filling a state, circles denote emission, squares denote absorption, and color identifies the dominant phonon branch. (e),(h) The instantaneous scattering flux resolved by branch, separated into emission and absorption, plotted on a shared logarithmic axis. (f),(i) The corresponding energy distribution $F(E)$. The population funnels from $\boldsymbol{\Gamma}$ to the $\mathbf{K}$ and $\mathbf{K^{\prime}}$ valley: about 45\% by $30$~fs and $90\%$ by 100 fs and remains there. The flux drops by three orders of magnitude between these two times -- not due to particle decay, as total number is conserved, but because the population settles into states where few scattering channels remain open.}
  \label{fgr:Fig_4}
\end{figure*}
The phonon-assisted PLE shown in Fig.~\ref{fgr:Fig_3}(a) confirms this prediction. This emission is carried entirely by the $E_1''$ mode, with all other branches vanishing to numerical precision (See Table~VII in SM~\cite{Supplemental}), and it produces a single satellite about 16~meV below the dark exciton (shaded region), matching the energy of this phonon at $\textbf{K}$.\\ 
\noindent \textit{Linewidth and rt-BTE---} Symmetry bounds what is possible, but it does not decide what happens. To separate its role from coupling strength and scattering kinematics, we treat the dynamics in two complementary limits. Following Chen et. al. \cite{Chen2020} the \textit{exc-ph} linewidth gives the out-scattering rate of state $|\lambda\mathbf{Q}\rangle$,
\begin{multline}\label{eq:fgr}
\gamma_{\lambda\mathbf{Q}}(T) = 2\pi \sum_{\beta\nu} \int_{\mathrm{BZ}}
  \frac{A_{\mathrm{uc}}\, d^2\mathbf{q}}{(2\pi)^2}
  \left|\mathcal{G}_{\beta\lambda\nu}(\mathbf{Q},\mathbf{q})\right|^2 \\
  \times \Big[ (N_{\nu\mathbf{q}}+1)\,
    \delta\!\left(E_{\lambda\mathbf{Q}} - E_{\beta\mathbf{Q}+\mathbf{q}}
                  - \hbar\omega_{\nu\mathbf{q}}\right) \\
  + N_{\nu\mathbf{q}}\,
    \delta\!\left(E_{\lambda\mathbf{Q}} - E_{\beta\mathbf{Q}+\mathbf{q}}
                  + \hbar\omega_{\nu\mathbf{q}}\right) \Big],
\end{multline}
where $\mathcal{G}_{\beta\lambda\nu}$ is the \textit{exc-ph} matrix element between excitons $\lambda$ and $\beta$ via phonon $\nu$, $N_{\nu\mathbf{q}}$ is the Bose--Einstein occupation at temperature $T$, and the stimulated factor of the final exciton state is neglected. The \textit{rt}-BTE, by contrast, follows the full exciton population $F_i(t)$, with $i \equiv (\lambda,\mathbf{Q})$, through the rate matrix $P_{ij}$ for scattering from $i$ to $j$, with phonons held at equilibrium~\cite{Chen2022,Rosati2020Cascade}
\begin{equation}\label{eq:bte}
\frac{dF_i}{dt} = -\sum_j \Big[ P_{ij} F_i (1 + F_j) - P_{ji} F_j (1 + F_i) \Big].
\end{equation}
\noindent The factors $(1+F)$ account for the bosonic statistics of excitons and replace the $(1-f)$ factors that encode Pauli blocking~\cite{Chen2022, Shah1996} in the corresponding \textit{el-ph} Boltzmann equation. With identical meshes, couplings, and delta-function integration, the out-scattering rate $\sum_j P_{ij}$ reproduces $\gamma_{\lambda\mathbf{Q}}/\hbar$ to within $0.1\%$. The two limits nonetheless answer different questions. At the dark minimum at $\textbf{K}$ shown in Fig.~\ref{fgr:Fig_3}(a), only phonon absorption contributes to $\gamma$. The excitons it removes are returned by phonon emission, so the state is broadened without losing population.\\
Although the two limits agree on the rates, the linewidth and the relaxation are mediated by different phonon branches. At 300~K, the lowest bright exciton at $\mathbf{Q}=\mathbf{0}$ has a linewidth of 151~meV, corresponding to a lifetime of about 4~fs shown in Fig.~\ref{fgr:Fig_3}(b)--(d). Of this width, absorption contributes to 43\%. 44\% of it comes from the lowest acoustic branch, and the rest from branches below 25~meV. The polar optical branches carry \textit{exc-ph} matrix elements an order of magnitude larger than any other mode, yet contribute nothing: they are even, and every final state within their energetic reach lies in the odd domain.\\
Temperature separates emission from absorption. Emission persists as $T\to0$ through the factor $N_\nu+1$, while absorption vanishes with $N_\nu$. At 5~K absorption is thermally frozen out, hence the linewidth of the bright state, about 37~meV, comes almost entirely from spontaneous phonon emission. Our broadening-free integration keeps this separation exact; a phenomenological Lorentzian broadening would blur it by admitting energy-nonconserving transitions. Resolved across the BZ as demonstrated in Fig.~\ref{fgr:Fig_3}(a),(e)--(f), the linewidth follows the available exciton space for phonon emission: it is smallest at the band minima, and grows with energy above them. Because lifetime is $\tau=\hbar/\gamma$, the narrowest states are also the longest lived. This observation is fully in accordance with the Heisenberg's uncertainty principle. At the minimum at $\textbf{K}$, no state lies below, and the width comes from quasi-elastic absorption of LA phonons of $\sim$1.5~meV within the valley, which leads to the lowest scattering rates. It rises linearly above 20~K as these phonons are thermally populated, and reaches about 10~meV at 77~K. Population, however, remains trapped in the valley for a much longer time than the lifetime inferred from the linewidth, so the linewidth at $\mathbf{K}$ primarily measures dephasing rather than population depletion. Higher-lying states instead broaden with energy as the exciton space for emission grows; the fourth exciton band at $\mathbf{Q=K}$ reaches 707~meV.\\
\noindent Figure~\ref{fgr:Fig_4} shows time-snapshots of the relaxation at $t=1$~fs, 30~fs, and after thermalization. After optical injection at $\mathbf{Q=0}$, with an in-plane light, weighted by the oscillator strengths at 300~K, the evolution of the exciton population across its BZ is shown in Fig.~\ref{fgr:Fig_4}(a)--(c). As shown in Fig.~\ref{fgr:Fig_4}(b)-(c), within 30~fs, about 45\% of the population reaches the $\mathbf{K}$ and $\mathbf{K}'$ valleys, while by thermalization about 90\% has accumulated there. At $t=30$~fs, the population projected onto the exciton dispersion is shown in Fig.~\ref{fgr:Fig_4}(d), where circles denote emission, squares denote absorption, and the color identifies the dominant phonon branch. Relaxation within the $\sigma_h$-even domain is blocked because the bright state is already at its lowest point. Instead, the odd ZA branch, which supplies the required momentum at low energy, carries the population across the parity boundary into the odd domain. Once there, even phonons take over: the fourth branch, in its in-plane form, carries the population down the odd band toward the $\mathbf{K}$ and $\mathbf{K}'$ valleys, while the polar optical branches open relaxation pathways with energy drops of about 100~meV. The corresponding branch-resolved scattering flux is shown in Fig.~\ref{fgr:Fig_4}(e), with branches 4 and 6 carrying the largest flux. The associated energy distribution in Fig.~\ref{fgr:Fig_4}(f) shows that the scattering remains concentrated in the energy range accessed during the cascade. Thus, the polar optical branches can dominate the relaxation flux even though they contribute nothing to the linewidth of the bright state. The \textit{rt}-BTE contains no decay channel, so the subsequent reduction of the flux reflects the population settling into states from which few scattering channels remain kinematically open.\\
\noindent After thermalization, the population distribution projected onto the exciton dispersion is shown in Fig.~\ref{fgr:Fig_4}(g). The state at $\mathbf{K}$ acts as a bottleneck because it cannot emit a phonon. The branch-resolved equilibrium flux in Fig.~\ref{fgr:Fig_4}(h) is therefore nearly three orders of magnitude smaller than during the initial cascade. The equilibrium energy distribution in Fig.~\ref{fgr:Fig_4}(i) is sustained by phonon absorption from states surrounding the minimum, as required by detailed balance. Although the total equilibrium flux is summed over all initial states, it retains an emission component because states above the minimum can still emit phonons. The resulting distribution has a mean energy of about 27~meV above the minimum, in agreement with the Boltzmann value of 27~meV evaluated over the same set of states.\\ 
\noindent \textit{Conclusion}---The horizontal mirror plane $\sigma_h$ gives every electron, phonon, and exciton state a parity that is exact at every wavevector, and this is what makes
it indispensable for understanding the real time \textit{exc-ph} dynamics. Branch indices mix character, and irreducible representations change from one little group to the next, but the $\sigma_h$ parity holds across
the whole zone. It divides the lowest exciton band into an even domain around $\mathbf{\Gamma}$ and an odd one containing the dark minimum at $\mathbf{K}$. In effect, the \textit{exc-ph} coupling reduces to a single rule, $p_\beta p_\nu p_\lambda = +1$. We used it in three ways. As a selection rule for emission, as an explanation for the absence of polar-optical contributions to the bright-state linewidth, and as a map of relaxation dynamics. Parity determines which channels exist; coupling strength determines how much each allowed channel carries. Since $\sigma_h$ is a symmetry of every $D_{3h}$ monolayer, the same analysis applies broadly to systems such as \textit{h}BN and GaN, where the mirror parities of the band-edge states determine the parity of the lowest $s$-like exciton: even and bright, as in \textit{h}BN, or odd and dark, as in GaN.\\  
\noindent \textit{Acknowledgements---}The author SB thank Anusandhan National Research Foundation (ANRF), India, under grant number CRG/2023/000476. This work is carried out using computational resources at PARAM Rudra 20 PF HPC System provided by Centre for development of advanced computing (C-DAC) under the National Supercomputing Mission (NSM), India, jointly steered by MeitY and DST.\\
\noindent \textit{Data availability---} The input files for this work are given in the SM~\cite{Supplemental} and the codes~\cite{xphd2026} developed for this work are available publicly.
\bibliography{prb}
\end{document}